\documentclass[acmtog]{acmart}
\acmSubmissionID{1332}
\usepackage{stfloats}
\usepackage{colortbl}
\AtBeginDocument{%
  }

\setcopyright{cc}
\setcctype{by}
\acmJournal{TOG}
\acmYear{2025}
\acmVolume{44}
\acmNumber{4}
\acmArticle{}
\acmMonth{8} 
\acmPrice{}
\acmDOI{10.1145/3730935}

\begin{document}

\definecolor{best}{rgb}{1.0, 1.0, 0.56}

\title{WishGI: Lightweight Static Global Illumination Baking via Spherical Harmonics Fitting}

\author{Junke Zhu}
\email{junkezhu@mail.ustc.edu.cn}
\orcid{0009-0005-7778-9614}
\affiliation{%
  \institution{University of Science and Technology of China and Tencent Technology}
  \country{China}
}

\author{Zehan Wu}
\email{zehanwu@tencent.com}
\orcid{0009-0006-2241-9816}
\affiliation{%
  \institution{Tencent Technology}
  \country{China}}
\email{zehanwu@tencent.com}

\author{Qixing Zhang}
\email{qixing@ustc.edu.cn}
\orcid{0000-0002-8784-8674}
\affiliation{%
 \institution{University of Science and Technology of China}
 \country{China}}
\email{qixing@ustc.edu.cn}

\author{Cheng Liao}
\email{chengliao@tencent.com}
\orcid{0009-0003-6540-3150}
\authornote{Corresponding authors.}
\affiliation{%
  \institution{Tencent Technology}
  \country{China}}
\email{chengliao@tencent.com}

\author{Zhangjin Huang}
\email{zhuang@ustc.edu.cn}
\orcid{0000-0003-1475-8894}
\authornotemark[1]
\affiliation{%
  \institution{University of Science and Technology of China}
  \country{China}}
\email{zhuang@ustc.edu.cn}


\begin{abstract}

Global illumination combines direct and indirect lighting to create realistic lighting effects, bringing virtual scenes closer to reality. Static global illumination is a crucial component of virtual scene rendering, leveraging precomputation and baking techniques to significantly reduce runtime computational costs. 
Unfortunately, many existing works prioritize visual quality by relying on extensive texture storage and massive pixel-level texture sampling, leading to large performance overhead. In this paper, we introduce an illumination reconstruction method that effectively reduces sampling in fragment shader and avoids additional render passes, making it well-suited for low-end platforms.
To achieve high-quality global illumination with reduced memory usage, we adopt a spherical harmonics fitting approach for baking effective illumination information and propose an inverse probe distribution method that generates unique probe associations for each mesh. This association, which can be generated offline in the local space, ensures consistent lighting quality across all instances of the same mesh.
As a consequence, our method delivers highly competitive lighting effects while using only approximately 5\% of the memory required by mainstream industry techniques.
\end{abstract}


\begin{teaserfigure}
  \includegraphics[width=\textwidth]{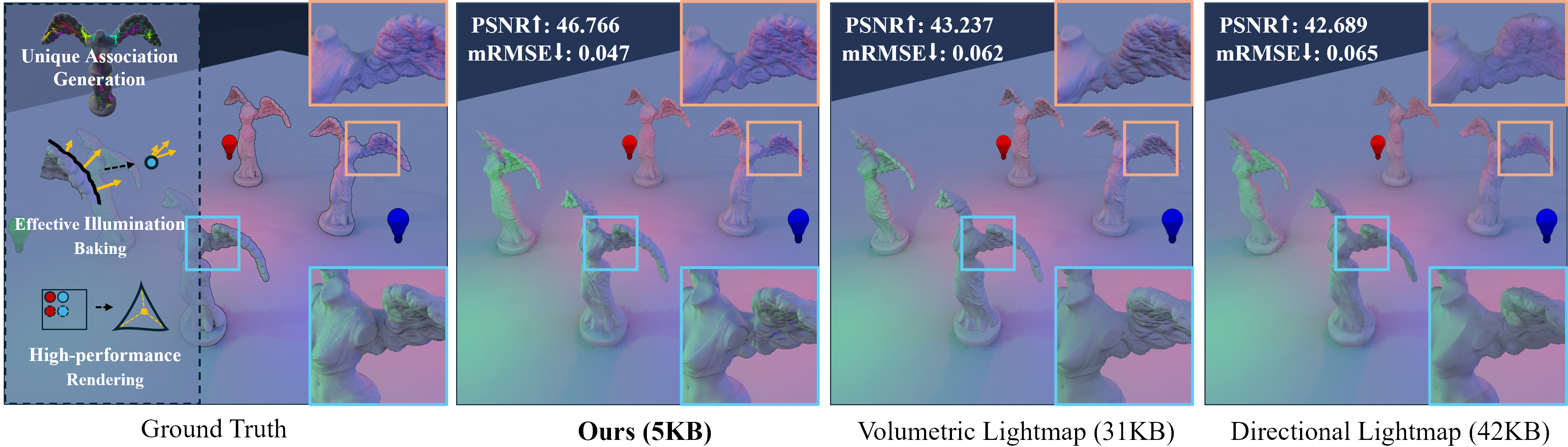}
  \caption{We present a novel method for static global illumination baking tailored for low-end platforms. Our work significantly reduces memory usage and ensures high runtime efficiency, while avoiding additional render passes, making it highly compatible with forward rendering. This method enables high-quality static global illumination on low-end devices, demonstrating clear advantages over mainstream industry methods.}
  \Description{}
  \label{fig:teaser}
\end{teaserfigure}


\maketitle

\section{Introduction}
Global illumination (GI) refers to the process of simulating all light interactions within a scene, including both direct and indirect lighting \cite{sota}. High-quality global illumination techniques can make virtual scenes more realistic, providing users with an enhanced experience. In most video games, static objects make up the majority of the scene. By leveraging precomputation and baking techniques, runtime computational costs can be significantly reduced. Therefore, static global illumination baking techniques \cite{intro1,intro2,intro3} have attracted a great deal of attention.

Among various static global illumination techniques, we focus on methods tailored for low-end platforms. Due to hardware limitations, these platforms often face a great challenge of memory and performance.
Existing probe-based works\cite{epic_vlm,probe1,probe2,NeuralLightGrid}
require storing a large number of spherical harmonics in the scene and rely on massive pixel-level interpolation during rendering, posing significant challenges to low-end platforms in terms of storage and runtime performance. 
Epic Games \cite{epic_ilc} just distributes probes on the surface of objects and reduces runtime overhead through per-component interpolation, making it feasible for mobile platforms. However, its lighting quality deteriorates when applied to structurally complex objects. 
Sloan et al. \cite{lightmap} calculate illumination based on the surface of the object and simplify the storage into light intensity and dominant light direction, which becomes the most widely adopted methods in industry. However, it introduces visual imperfections in low-end platforms due to limited precision and compression artifacts. Moreover, the gaps caused by the use of UV mapping result in wasted memory space.

In this paper, we introduce a novel illumination reconstruction method, which supports Level of Detail (LOD), effectively reduces sampling in fragment shader and avoids additional render passes. Therefore, this method is well-suited for forward rendering (a rendering technique widely used in Tile-based Deferred Rendering  GPUs). To achieve high-quality global illumination with minimal memory usage, we adopt a spherical harmonics fitting approach that efficiently bakes effective illumination information to ensure accuracy. Since existing probe-based works typically place fixed probes in the scene leading to redundancy in spherical harmonics when applied to baking static objects, we propose an inverse probe distribution method that generates unique probe associations for each mesh in local space through optimization. This association, which can be generated offline, ensures consistent lighting quality across all instances of the same mesh. We demonstrate our work through illumination quality and performance statistics.

The major contributions of this work are as follows.
\begin{itemize}
\item We propose an illumination reconstruction method that effectively reduces sampling in fragment shader and avoids additional render passes, ensuring exceptional runtime performance.
\item We propose an illumination baking method that bakes effective lighting information via spherical harmonics fitting, improving the utilization of effective information and achieving higher precision.
\item We propose an inverse probe distribution method that generates unique probe associations for each mesh in the local space to reduce spherical harmonics redundancy. This association can be generated offline and ensures consistent lighting quality across all instances of the same mesh.

\end{itemize}

\section{RELATED WORK}

\hspace{0.6\parindent}
\textit{Hemispherical illumination.}
The illumination of objects in the real world usually varies with changes in surface normals. To achieve detailed lighting effects, illumination information must be recorded in the hemispherical space defined by the geometric normals. Thus, spherical harmonics (SH) have been adopted as a medium for representing illumination since earlier works \cite{564,257}. \citet{PRT} also introduced the SH-based concept of precomputed radiance transfer, and subsequent works \cite{1653, 10.1145/3320284,intro3} achieved remarkable visual results. However, this technology is far from being supported on low-end platforms due to its significant memory requirements. Simpler representations, such as AHD (Ambient/Highlight/Direction) \cite{809,998,806}, significantly reduce storage requirements compared to SH, but AHD's nonlinear nature makes linear interpolation of individual components mathematically inaccurate. Environment cube-based representations \cite{808,766,533} also provide a good way to represent illumination by storing limited surface data and interpolating to compute colors for arbitrary directions, but they still lag behind spherical harmonics in terms of quality.

Our work adopts spherical harmonics to encode hemispherical illumination information at different sample points, preserving more lighting details. Their linearity ensures the feasibility and accuracy of surface interpolation, making the entire baking process differentiable and facilitating subsequent optimization.

\textit{Global illumination baking.}
2D lightmap is one of the most commonly used methods for storing precomputed lighting. Research in this area includes parameterization \cite{1057}, packing and compression \cite{1036,10.1145/3617683}, and seam handling \cite{1058,1467}. \citet{lightmap} proposed Directional Lightmap, which stores lighting intensity and the direction of the strongest incoming light. While lightmaps strike a good balance between visual quality and runtime performance, visual imperfections are often introduced on low-end platforms due to limited resolution and compression artifacts. Moreover, the gaps caused by the use of UV mapping result in wasted memory usage. \citet{10885008} attempted to store incoming radiance information in point clouds, yet this approach still incurred considerable storage demands, posing challenges for complex scenes. The lightweight probe-based method Indirect Light Caching \cite{epic_ilc} is a common solution on mobile platforms, performing per-component spherical harmonic interpolation. However, this approach often results in poor lighting quality for large objects or those with complex structures.

The other probe-based works typically require pixel-level sampling for rendering. \citet{316} interpolated values through tetrahedral vertices; Volumetric Lightmap \cite{epic_vlm}, implemented by Epic Games, performs trilinear interpolation per pixel using nine 3D textures; \citet{ddgi} sample each probe in the eight-probe cage using the surface normal in world space to calculate the pixel color, etc. The large amount of pixel-level sampling place a significant burden on fragment shader, posing a considerable challenge for low-end platforms. Meanwhile, these works typically require storing large amounts of data to ensure the accuracy of spherical harmonics. Other recent works introducing neural network into scene representation \cite{nenv,Granskog,Sitzmann} can achieve good visual effects, but it is not feasible for low-end devices due to both memory usage and runtime efficiency.


\textit{Probe distribution.}
Probes can be defined in various forms.
In the field of illumination baking, probes are viewed as Spherical Harmonics (SH) expansions \cite{PRT}. The probe distribution strategy is a crucial part of SH-based techniques. Poorly placed probes can lead to interpolation artifacts, such as light leakage. Probe positions can be determined either manually \cite{316} or automatically \cite{809,1812}. \citet{NeuralLightGrid} combined light field information with pre-trained neural networks to distribute the probes, aiming to reduce light leakage. Existing works typically distribute probes by analyzing the lighting conditions of the scene, which often results in spherical harmonics redundancy. Moreover, a reasonable probe distribution must be explored before each baking process.

On the other hand, \citet{currius2020spherical} define probes as Spherical Gaussians, treating probe distribution as an optimization problem. Recently, a series of works \cite{3dgs,lu2024scaffold,NeLF-Pro} based on Spherical Gaussians optimization have achieved remarkable results in novel view synthesis. Approaching probe distribution as an optimization task and performing an inverse probe distribution often yields better results with fewer probes. But performing the optimization for the entire scene before baking leads to lengthy lighting build times due to the vast amount of scene information (geometric \& lighting data), making it impractical. In this paper, we perform an inverse probe distribution in the local space for each mesh, which can be conducted offline once after mesh import. This information is then embedded into the mesh, avoiding the need for probe distribution before light baking.
 

\section{METHOD}
\subsection{Problem and formulation}

\begin{figure}[h]
  \includegraphics[width=\linewidth]{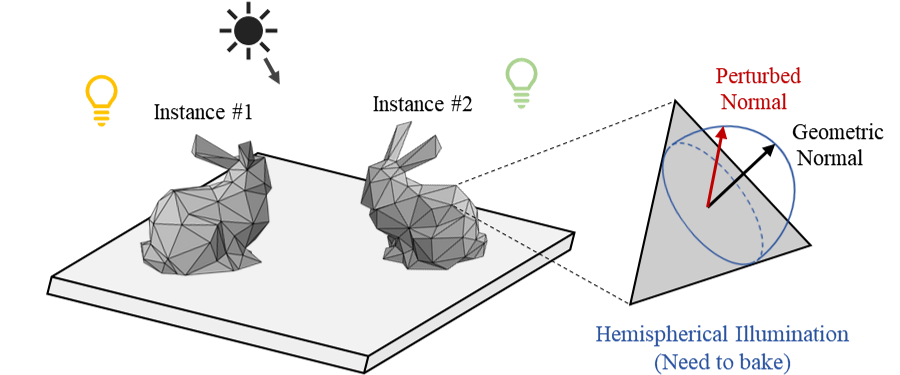}
  \caption{Normal maps perturb the geometric normals in tangent space, and illumination is calculated based on the perturbed normals. Thus, hemispherical illumination is the target we aim to bake. Additionally, since multiple instances of the same model can exist in a scene, operations in the local space must be applicable to all instances.}
  \label{fig:problem}
\end{figure}

\hspace{0.6\parindent}
\textit{Problem statement.}
Our work based on probes (Spherical Harmonics) aims to achieve high-quality static global illumination baking with extremely low memory usage. During illumination reconstruction, our method should achieve excellent runtime performance, supporting low-end platforms, such as mobile devices.
To preserve rich lighting details even after normal map perturbations, we need to bake illumination information for the hemispherical space $\Omega$ corresponding to the geometric normal, as shown in Figure \ref{fig:problem}.

\textit{Formulation.}
The input data is a triangular mesh within a scene, composed of vertices $\mathbf{v}_1, \mathbf{v}_2, \dots, \mathbf{v}_{N_v}$. We assume the existence of an integrable function $f : S \to \mathbb{R}^3$  that maps each point on the surface $S$ of the mesh to its corresponding hemispherical space illumination.

In addition to designing a high-performance illumination reconstruction scheme that supports low-end platforms, the key question is how to design an efficient light baking method that uses minimal storage to achieve high-quality static global illumination. Thus, the problem can be formulated as an optimization problem, aiming to minimize the difference between the reconstructed global illumination $\hat{f}(S)$ and the actual illumination $f(S)$ across the entire surface $S$:
\begin{equation}
\label{func:goal}
    \min \ E_{\textit{loss}}\left( \hat{f}(\mathcal{A}; S), f(S)\right),
\end{equation}
where $\hat{f}$ is the reconstructed illumination with the association parameters $\mathcal{A}$. $E_\textit{loss}$ is the loss function that measures the similarity between the reconstructed and actual illumination. $\mathcal{A}$ represents the result of probe distribution in our work generated in local space for each mesh, specifically the association between vertices and probes through their indices and weights.

\textit{Challenges.} 
First, previous works \cite{316,epic_vlm} based on spherical harmonics perform massive texture sampling within the fragment shader, resulting in significant runtime overhead that is unacceptable for low-end platforms. We need to design a differentiable illumination reconstruction model that achieves high-quality global illumination while maintaining exceptional runtime efficiency. Second, probe-based methods typically require substantial memory usage and often suffer from visual issues such as light leakage and ringing artifacts \cite{stupidsh}. Our approach should minimize memory consumption while ensuring visual fidelity. Third, reuse of meshes is very common in practical scenarios. The association parameters $\mathcal{A}$ generated in local space for each mesh should deliver consistent and reliable results across different instances of the same mesh. In this way, the association can be constructed offline and embedded within the mesh, avoiding exploring probe distribution in the scene before light building.

\subsection{High-performance illumination reconstruction}
Existing approaches \cite{epic_vlm,sh1,intro1} leverage probes (Spherical Harmonics, SH) to store spherical lighting information, achieving excellent illumination effects. However, these methods place a significant load on the fragment shader due to massive pixel-level sampling, which poses performance challenges on low-end platforms. 
To address the limitation, we propose a vertex-based illumination model that reconstructs illumination using minimal computation, ensuring excellent runtime performance. Additionally, we encode the coefficients to reduce the number of samples while seamlessly supporting Level of Detail(LOD).

\begin{figure}[h]
  \includegraphics[width=\linewidth]{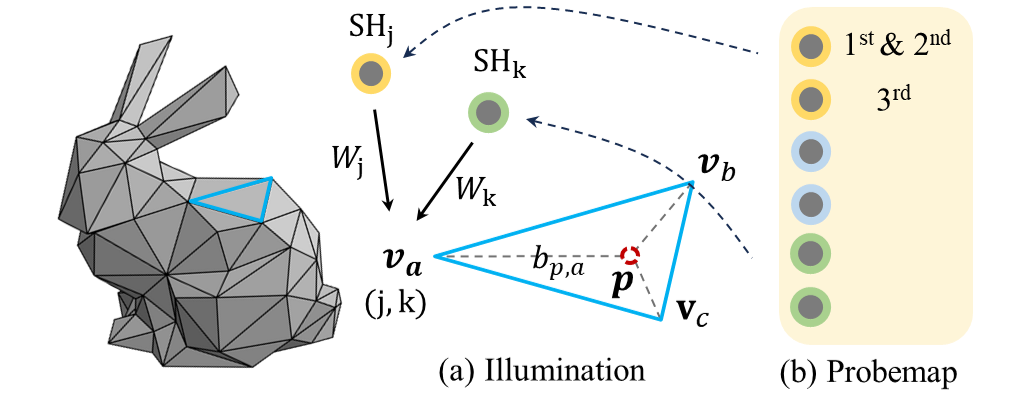}
  \caption{(a) Retrieve values from the probemap using indices and weights, then decode and combine for illumination calculation. (b) All probe data for the scene is stored seamlessly in a single probemap.}
  \label{fig:highperformance}
\end{figure}

\textit{Illumination reconstruction function.} We treat multiple sets of spherical harmonics (Probes) as higher-dimensional basis functions. The spherical harmonic values at each vertex, denoted as $SH_{\mathbf{v}_i}$, are represented as a linear combination of the probes $SH_i$. The association parameters $\mathcal{A}$ directly record the relationship between vertices and probes in the form of indices $j,k,...,n$ and weights $W_j,W_k,...,W_n$:
\begin{equation}
     SH_{\mathbf{v}_i} = W_j SH_j + W_k SH_k+... + W_n SH_n.
\end{equation}

For each point $\mathbf{p} \in S$, its spherical harmonic values are interpolated using the barycentric coordinates $\mathbf{b} = (b_{p,a},b_{p,b},b_{p,c})$ of the three vertices $\mathbf{v}_a$, $\mathbf{v}_b$, $\mathbf{v}_c$ of its containing triangle. In this way, as shown in Figure \ref{fig:highperformance}(a), we define the illumination reconstruction formula for point $\mathbf{p}$ as:
\begin{equation}
\label{func:reconstruction}
    \hat{f}(\mathcal{A};\mathbf{p}) = \frac{1}{\pi}\sum_{l=0}^{\infty} \sum_{m=-l}^{l} (b_{p,a} SH_{\mathbf{v}_a} + b_{p,b} SH_{\mathbf{v}_b} + b_{p,c} SH_{\mathbf{v}_c})_{l}^{m} T(Y_{l}^{m}(\mathbf{d})),
\end{equation}
where $Y_{l}^{m}(\mathbf{d})$ represents a spherical harmonics basis function for a given degree $l$ and order $m$, evaluated at the direction $\mathbf{d}$, and $T$ performs the convolution of a SH with the circularly symmetric function \cite{stupidsh}.

The formula above enables efficient illumination reconstruction with minimal computational complexity and places more emphasis on the vertex compared to other works \cite{epic_vlm,ddgi}. Moreover, this computation can be performed directly in the base pass without any additional render pass, making it perfectly suited for forward rendering and well adapted to Tiled-Based Deferred Rendering (TBDR) on mobile chips.

\begin{figure}[h]
  \includegraphics[width=0.9\linewidth]{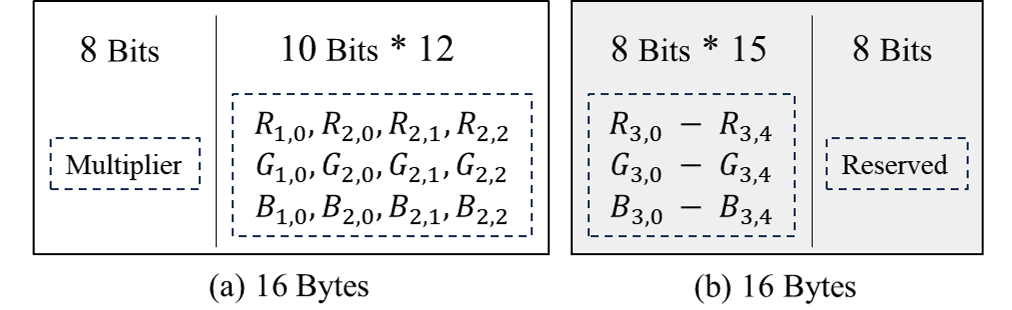}
  \caption{(a) The first pixel (16 bytes) stores the Multiplier and the coefficients for the first and second orders. (b) The second pixel (16 bytes) stores the coefficients for the third order, leaving 8 bits reserved.}
  \label{fig:compress}
\end{figure}

\textit{Coefficients encoding \& storage.}
Humans are generally more sensitive to the color channel with the highest brightness \cite{wallace1991jpeg}. Therefore, we store the maximum absolute value among the three color channels of the spherical harmonic coefficients as a multiplier, using one byte to ensure accuracy. The other coefficients are normalized using this value, allowing us to uniformly control the precision of the spherical harmonic coefficients. Considering that most hardware platforms support high-precision texture formats like $R32G32B32A32$, we use twelve 10-bits to store the coefficients of the first- and second-order spherical harmonics, while the third-order coefficients are stored using fifteen 8-bits, as shown in Figure \ref{fig:compress}.

Note that we only need a single multiplication operation to decode the data. This further optimizes the sampling frequency, enabling support for Level of Detail (LOD): distant objects use second-order spherical harmonics with a single texture sample, while closer objects utilize third-order spherical harmonics requiring two texture samples. Additionally, as shown in Figure \ref{fig:highperformance}(b), our approach eliminates both the gaps in texture \cite{lightmap} and the requirement to store multiple textures \cite{epic_vlm,PRT}. All probe data for the entire scene can be seamlessly stored in a single probemap in our work, supporting maximum instance draw and reducing draw call overhead.

\subsection{Effective illumination baking}

Now, we proceed to bake the illumination information into probes. The baked information will be output through the aforementioned reconstruction function. Since we focus exclusively on the illumination of static objects, unlike existing methods \cite{epic_vlm,probe1}, we do not need to bake the complete environmental lighting. Instead, our goal is the precise illumination representation of the object's surface. Therefore, we only fit the illumination information along the effective directions of valid sampling points on the surface $S$. For the illumination across the entire hemisphere, we prioritize ensuring accuracy in the direction of the mesh's normal. Thus, we define the weight for each direction as $w(\mathbf{d}) = \max(0, \cos(\mathbf{d}, \mathbf{n}))$, where $\mathbf{d}$ is the spherical sampling direction and $\mathbf{n}$ is the surface normal. We define the loss function $E_{\textit{light}}$ as follows:
\begin{equation}
    E_{\textit{light}} = \int_{S} \left( \int_{\Omega} w(\mathbf{d}) \left( \hat{f}(\mathcal{A};\mathbf{p}, \mathbf{d}) - f(\mathbf{p}, \mathbf{d}) \right)^2 \, d\mathbf{d} \right) d\mathbf{p},
\end{equation}
where $f(\mathbf{p}, \mathbf{d})$ is the actual illumination of a valid direction $\mathbf{d}$ at sample point $\mathbf{p}$, and $\hat{f}(\mathcal{A}; \mathbf{p}, \mathbf{d})$ is the reconstructed illumination based on the association parameters $\mathcal{A}$. 

Considering that global illumination is typically low-frequency and smooth, we introduce a regularization term to ensure visual smoothness. The pixel value is calculated from the vertices using barycentric coordinates in Eqn. (\ref{func:reconstruction}). Therefore, we only need to constrain the gradient along the geometric normal direction $\mathbf{n}$ of the vertices $\mathbf{v}$ to ensure smoothness:
\begin{equation}
    E_{\textit{reg}} = \sum_{\mathbf{v} \in S}||\nabla \hat{f}(\mathcal{A}; \mathbf{v}, \mathbf{n})||^2.
\end{equation}

Inspired by Kavan et al. \cite{ao}, we adopt the edge-based approach to compute $\nabla \hat{f}$, which focuses on the gradients of adjacent triangles in a butterfly-shaped configuration. The calculation formula is as follows:
\begin{small}
\begin{equation}
   \sum_{(t,u)}(a_t+a_u)||\nabla \hat{f}_{|t} - \nabla \hat{f}_{|u}||^2 = \sum_{(t,u)}(a_t+a_u)\mathbf{L}^T \left(\mathbf{F}_t-\mathbf{F}_u\right)^T \left(\mathbf{F}_t-\mathbf{F}_u\right)\mathbf{L},
\end{equation}
\end{small}
where \((t,u)\) represents a pair of adjacent triangles, \(a_t\) and \(a_u\) are areas of the triangles, $\mathbf{L}$ is the matrix of vertex illumination values and $\mathbf{F_i}$ is a sparse matrix that $\mathbf{F}_i \mathbf{L} = \nabla \hat{f}_{|i}$. For simplicity, we will use the difference matrix \( \mathbf{D} \) to represent \( \sqrt{a_t+a_u}(\mathbf{F}_t - \mathbf{F}_u )\). 

Since the association parameters \( \mathcal{A} \) of the mesh to be baked are determined in this section, we use the weight matrix \( \mathbf{W} \) to represent the association between the vertices and probes. We further express the regularization term $E_\textit{reg}$ as:
\begin{equation}
    E_{\textit{reg}}= ||\mathbf{D} \cdot \mathbf{Y}' \cdot \mathbf{W} \cdot \mathbf{SH} ||^2,
\end{equation}
where $\mathbf{Y}'$ is the spherical harmonic basis at the vertices after applying the convolution function $T$ \cite{stupidsh}.

Our final loss function combines the lighting loss and the regularization term and the final optimization objective is defined as:
\begin{equation}
    \min \limits_{\mathbf{SH}} E_{\textit{loss}} = \min \limits_{\mathbf{SH}} \left( E_{\textit{light}} + \lambda E_{\textit{reg}}\right).
\end{equation}

Note that this optimization objective is differentiable with respect to $\mathbf{SH}$, we can find its global minimum by solving ${\partial E_{\textit{loss}}}/{\partial \mathbf{SH}} = 0$, which finally reduces to:
\begin{small} 
\begin{align}
    ((\mathbf{w} \cdot T(\mathbf{Y}) \cdot \mathbf{B} \cdot \mathbf{W})^T (\mathbf{w} \cdot T(\mathbf{Y}) \cdot \mathbf{B} &\cdot \mathbf{W}) + \lambda((\mathbf{Y}'\mathbf{W})^T \mathbf{D}^T \mathbf{D} (\mathbf{Y}'\mathbf{W})))\mathbf{SH} \nonumber \\
    &= (\mathbf{w} \cdot T(\mathbf{Y}) \cdot \mathbf{B} \cdot \mathbf{W})^T \cdot (\mathbf{w} \cdot \mathbf{I}) \label{eq:example}
\end{align}
\end{small}
where matrix $\mathbf{B}$ represents the barycentric coordinates, matrix $\mathbf{I}$ is the actual illumination in the effective directions of valid points, and $\mathbf{SH}$ is the matrix of all spherical harmonic coefficients (probe data), which is the target we are solving for. The problem of baking illumination information into probes is successfully reformulated as a classic linear regression problem. Detailed derivation can be found in the supplementary.

\begin{figure}[h]
  \includegraphics[width=\linewidth]{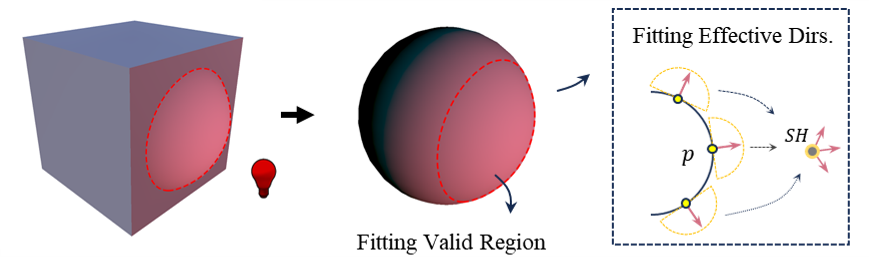}
  \caption{We fit the values for effective directions of valid sample points, allowing invalid regions or directions to take arbitrary values to ensure the precision.}
  \label{fig:valid}
\end{figure}

In this way, we can efficiently complete the illumination baking. Unlike previous works that compute spherical harmonic coefficients through SH projection \cite{10.1145/383259.383317} using uniform sampling in all directions, our method prioritizes the accuracy of illumination in the effective directions of valid points, as shown in Figure \ref{fig:valid}. This approach maximizes the utilization of effective illumination information while ensuring that irrelevant data do not impact the precision of hemispherical illumination within the normal space.

\subsection{Inverse probe distribution}
Now, let us perform the object-level probe distribution. Unlike conventional probe-based works \cite{epic_vlm, epic_ilc, PRT} treating probes as entities in the space, probes in our work serve solely as spherical harmonics for transferring illumination, without including any other information such as position, due to the direct association between vertices and probes. Given that a single mesh model has limited geometric details and the probes are solely related to the mesh rather than the entire scene, we treat the probe distribution as an optimization problem for the association $\mathcal{A}$ in local space.

\textit{Loss function.} Due to object reuse in a scene is common, we aim for all instances of the same mesh shared the unique association parameters \( \mathcal{A} \) can achieve consistent and high-quality results. The loss of evaluating the parameters \( \mathcal{A} \) is defined as:

\begin{equation}
\label{func:optimize}
    \min \limits_\mathcal{A} E_{\textit{loss}} = \min \limits_\mathcal{A} \sum_{i=1}^{N_{sce}} E_{\textit{light}}^{(i)}(\mathcal{A}),
\end{equation}
where $N_{sce}$ scenarios with different lighting conditions are constructed to evaluate the quality.

\textit{Geometric prior.} Light leakage is a common issue in probe-based methods. To mitigate this problem, we initialize probes based on geometric priors, which shortens optimization time as well. To adapt to our pipeline and ensure efficient illumination reconstruction, we utilize classical machine learning techniques rather than neural processes.

We first apply a K-Medoids clustering method \cite{km} on the valid sample points:
\begin{equation}
E = \sum_{i=1}^{K} \sum_{\mathbf{p} \in C_j} \text{dist}(\mathbf{p}, o_i), 
\end{equation}
where $E$ is the total cost, representing the sum of distances between each sample point $\mathbf{p}$ and its associated medoid $o_i$. $K$ denotes the number of medoids. $C_j$ is the cluster of points associated with the $j$th medoid. This approach positions cluster centers on the mesh surface, enabling efficient distance computation through sampled point connectivity as follows. This clustering occurs as a one-time operation during mesh import, with probes initialized based on the medoids.

\begin{figure}[h]
  \includegraphics[width=\linewidth]{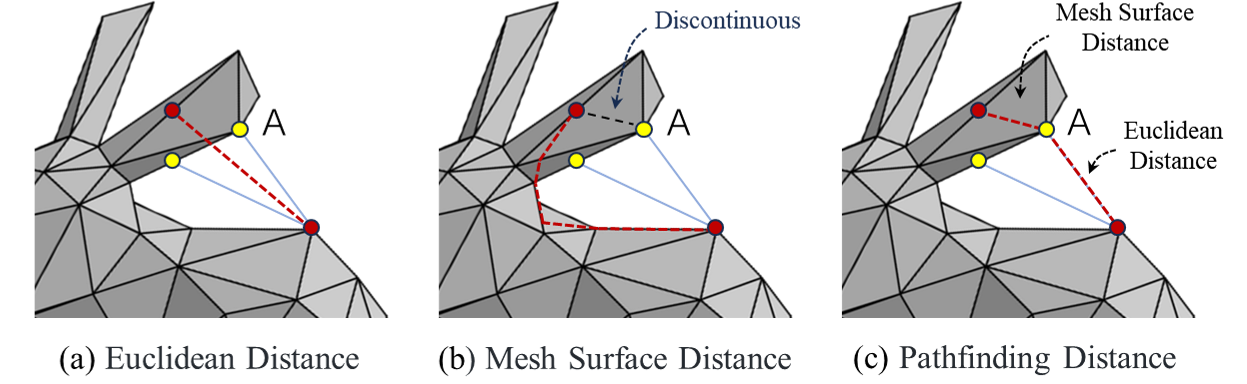}
  \caption{Red sampling points are not mutually visible. (a) Euclidean distance results in close sampling points, making separation difficult and causing light leakage; (b) Mesh surface distance causes discontinuities at the boundary between visible and non-visible points (point A); (c) Visible points use Euclidean distance for graph construction, non-visible points use pathfinding to increase occluded point distance, ensuring smooth transitions.}
  \label{fig:pathfinding}
\end{figure}

Inspired by Iwanicki et al. \cite{NeuralLightGrid}, we combined the heuristic-based approach of A* \cite{A*} with the exhaustive search nature of Dijkstra's algorithm to refine the distance metric. If two sample points are mutually visible, the Euclidean distance will be the distance between them. If they are not visible, they are considered disconnected. This process then constructs a graph connecting all the sample points. Finally, the distance between two non-visible points is calculated using a pathfinding algorithm among the sample points as shown in Figure \ref{fig:pathfinding}(c). In this way, we appropriately increased the distance for occluded sample points while maintaining continuity.

\begin{figure}[h]
  \includegraphics[width=0.9\linewidth]{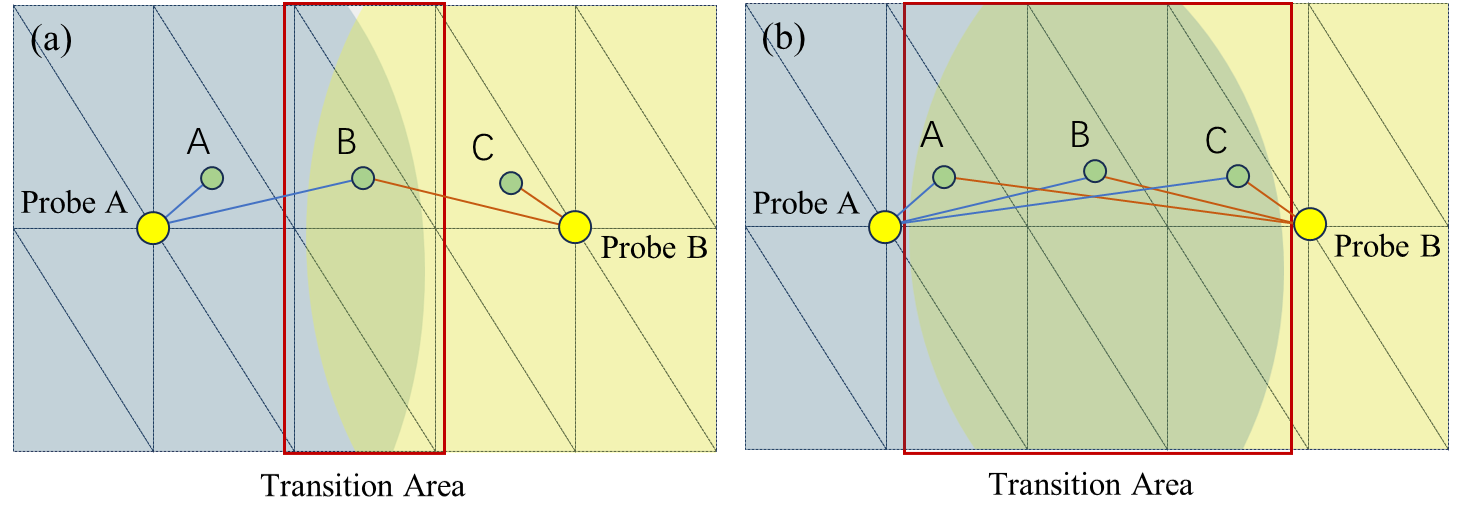}
  \caption{(a) Each sampling point is associated only with the nearest probe, resulting in transition area existing only in the edge triangles, which would lead to visual artifacts, especially when the mesh triangles are dense. (b) As each sampling point is associated with more probes, the transition area becomes larger, resulting in smoother outcomes.}
  \label{fig:oneprobe}
\end{figure}

After clustering, we calculate the distances to the nearest $n$ medoids (Probes) for each sample point, invert these distances, and normalize them to obtain the weights $w_{o_1}, w_{o_2}, \dots, w_{o_n}$. These weights are then transferred to the vertices using barycentric coordinates: 
\begin{equation}
W_{o_i} = \sum_{\mathbf{p} \in \{\mathbf{p}_1, \mathbf{p}_2, \ldots, \mathbf{p}_m\}} \left( w_{o_i} \cdot \textit{bary}(\mathbf{p}) \right)
\end{equation}
where $\mathbf{p}_1,\mathbf{p}_2 \ldots \mathbf{p}_m$ are  the sample points within one triangle, $W_{o_i}$ is the weight of a vertex for probe $o_i$, $w_{o_i}$ is the normalized weight of a sample point $\mathbf{p}$ for probe $o_i$ and $\textit{bary\((\mathbf{p})\)}$ means the barycentric coordinate at point $\mathbf{p}$. Each vertex needs to be assigned the top $n$ probes with the highest weights as geometric priors to avoid visual discontinuities in Figure \ref{fig:oneprobe}(a) when the mesh triangles are dense.

\textit{Optimization.}
Now, we have preliminarily initialized the association parameters \(\mathcal{A}\) between vertices and probes based on geometric prior. We can still use the weight matrix $\mathbf{W}$ to represent the associations, but now $\mathbf{W}$ is the objective that we need to optimize. 

We construct a standard scene with complex lighting, place the mesh into it, and simulate its illumination fitting under various complex scenarios by rotating the mesh. Essentially, this process encodes geometric information through lighting. The comprehensive loss defined in Eqn. (\ref{func:optimize}) is then calculated. This loss is differentiable with respect to the matrix $\mathbf{W}$, allowing us to further optimize it using gradient descent. In this way, we effectively mitigate light leakage and reduce the ringing artifacts of spherical harmonics.

Note that our work in this section is conducted entirely in local space after model import, effectively reducing probe redundancy, and ensuring consistent, high-quality results across all instances of the mesh. Meanwhile, the association can be generated offline and embedded into the mesh, avoiding the need to explore probe distribution in the scene.

\section{IMPLEMENTATION DETAILS}
\begin{figure}[h]
  \centering
  \includegraphics[width=\linewidth]{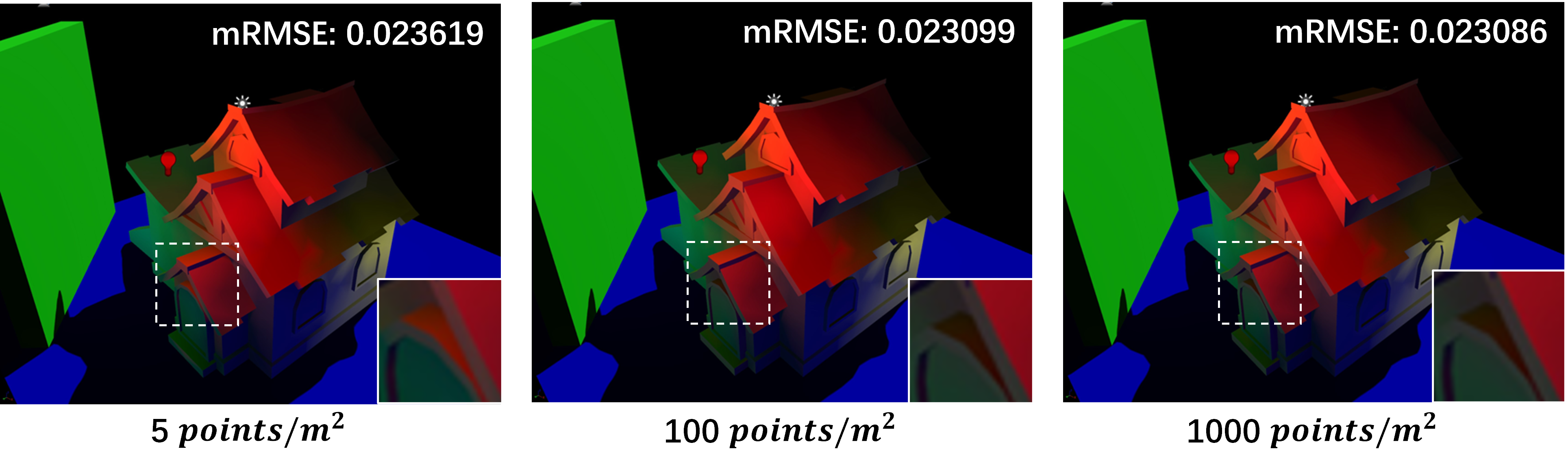}
  \caption{Visual artifacts are noticeable at lower sampling densities. Beyond 100 points/m\(^2\), further increasing the density yields limited improvements.}
  \label{fig:density}
\end{figure}

\hspace{0.6\parindent}
\textit{Sampling density.}
We use the Blue Noise Sampling method \cite{6197186} to perform uniform sampling on the surface of the mesh. We tested different sampling point densities and some of the results are shown in Figure \ref{fig:density}. Ultimately, we set 100 points per square meter as our standard.  

\textit{Probe associations per vertex.}
We associate each sampling point with the two probes. Thus, each vertex is associated with two probes. The association data for each vertex is stored as indices and weights in $R8G8B8A8$ format, with 8 bits allocated for the probe index and another 8 bits for the weight. Based on our tests, 256 probes are sufficient to represent even the most complex objects, with most requiring fewer than 30 probes for high-precision representation. The weights and indices are stored by replacing the memory allocated for UV2 at each vertex, introducing no additional memory overhead.

\textit{Empirical parameter setup.} 
During illumination baking, we set the smoothness regularization weight $\lambda$ to 0.1. We performed spherical sampling with 960 directions, fitting the illumination for effective directions that are on the same side as the surface normal. In the probe distribution, we built our code framework using LibTorch and employed the Adam optimizer with a learning rate decayed from 0.01 to 0.001, running for 400 iterations. To reduce the computational overhead during iterations, the probe distribution optimization used spherical sampling with 120 directions, fitting illumination for effective directions with angles less than 30 degrees relative to the surface normal.

\textit{Baking process time.}
Our work consists of two main components during the production process: probe distribution and light baking. We provide the timing details of our current implementation for reference. The time can vary significantly depending on the specific implementation and hardware setup. First, the time required for probe distribution for each mesh is typically within 2 minutes. Since this is a one-time task performed in the local space after mesh import, it does not affect the production process. Second, the baking time we recorded for scenarios as shown in Table \ref{tab:bakingtime}. The baking time of our work is comparable to that of the Directional Lightmap \cite{epic_lightmap} (the official implementation in Unreal Engine). Note that both of the aforementioned tasks are part of the production process and are conducted offline. The runtime performance of our work will be detailed in Section 5.

\begin{table}[h]
\caption{The baking time obtained from a single PC equipped with an i9-10980XE CPU, an RTX 3090 GPU, and 128 GB of RAM.}
\label{tab:bakingtime}
\resizebox{0.45\textwidth}{!}{
    \begin{tabular}{l|c|c|c|c}
    \toprule
     & \multicolumn{1}{c|}{Sun Temple} & \multicolumn{1}{c|}{Egypt} & \multicolumn{1}{c|}{Small Scene} & \multicolumn{1}{c}{Large Scene}\\ 
    \midrule
    Lightmap & 16\textit{mins} & 19\textit{mins} & 13\textit{mins} & 80\textit{mins}\\
    Ours     & 21\textit{mins} & 25\textit{mins} & 19\textit{mins} & 107\textit{mins} \\
    \bottomrule
    \end{tabular}
}
\end{table}

\section{RESULTS}
The implementation of this work was developed on Unreal Engine 4.27. We first demonstrate the effectiveness of our method on individual objects, followed by results of entire scenes. Next, we provide performance evaluations of our method on mobile devices.

\subsection{Illumination quality evaluation}

\hspace{0.6\parindent}
\textit{Ground truth.} 
We performed high-density sampling on the mesh surface $S$ and recorded the radiance values obtained from ray tracing for each sample point using third-order spherical harmonics, serving as the numerical ground truth for quantitative analysis. The visual reference ground truth is computed based on the directions perturbed by the normal map.

Using this ground truth as a standard reference, we compare our work with the current mainstream methods, Directional Lightmaps (Lightmap) \cite{lightmap} and Volumetric Lightmaps (VLM) \cite{epic_vlm}, which share the same source data.


\textit{Evaluation metrics.} 
We defined a multi-directional RMSE (\textit{mRMSE}) to assess the results. This metric evaluates the quality of hemispherical illumination within the hemisphere defined by the geometric normal, thereby assessing the lighting quality after perturbations caused by normal maps. The calculation formula is as follows:
\begin{small}
\begin{equation}
\textit{mRMSE} = \sqrt{\sum_{i=1}^{N_s} \left( \frac{\textit{area}_i}{\textit{Area}} \int_{\Omega} \frac{\cos(\theta)}{\textit{Cos}} \left( f(\mathbf{d}) - \hat{f}(\mathbf{d}) \right)^2 \, d\mathbf{d} \right)}
\end{equation}
\end{small}
where \( N_s \) is the number of sampling points, \( \frac{\textit{area}_i}{\textit{Area}} \) and \( \frac{\cos(\theta)}{\textit{Cos}} \) respectively represent the normalizations over the area and hemispherical angles, \( f(\mathbf{d}) \) and \( \hat{f}(\mathbf{d}) \) are the ground truth and the illumination reconstructed. Lower \textit{mRMSE} values indicate higher accuracy in the global illumination baking. The details can be found in the supplementary. 

Note that we directly compare the differences between the reconstructed values and the numerical ground truth, rather than evaluating the final rendered image. This metric transcends specific image spaces, allowing for a comprehensive evaluation across the entire object. Additionally, it eliminates the influence of other rendering factors, enabling a more precise and efficient comparison.

\begin{figure*}[p]
  \includegraphics[width=\textwidth]{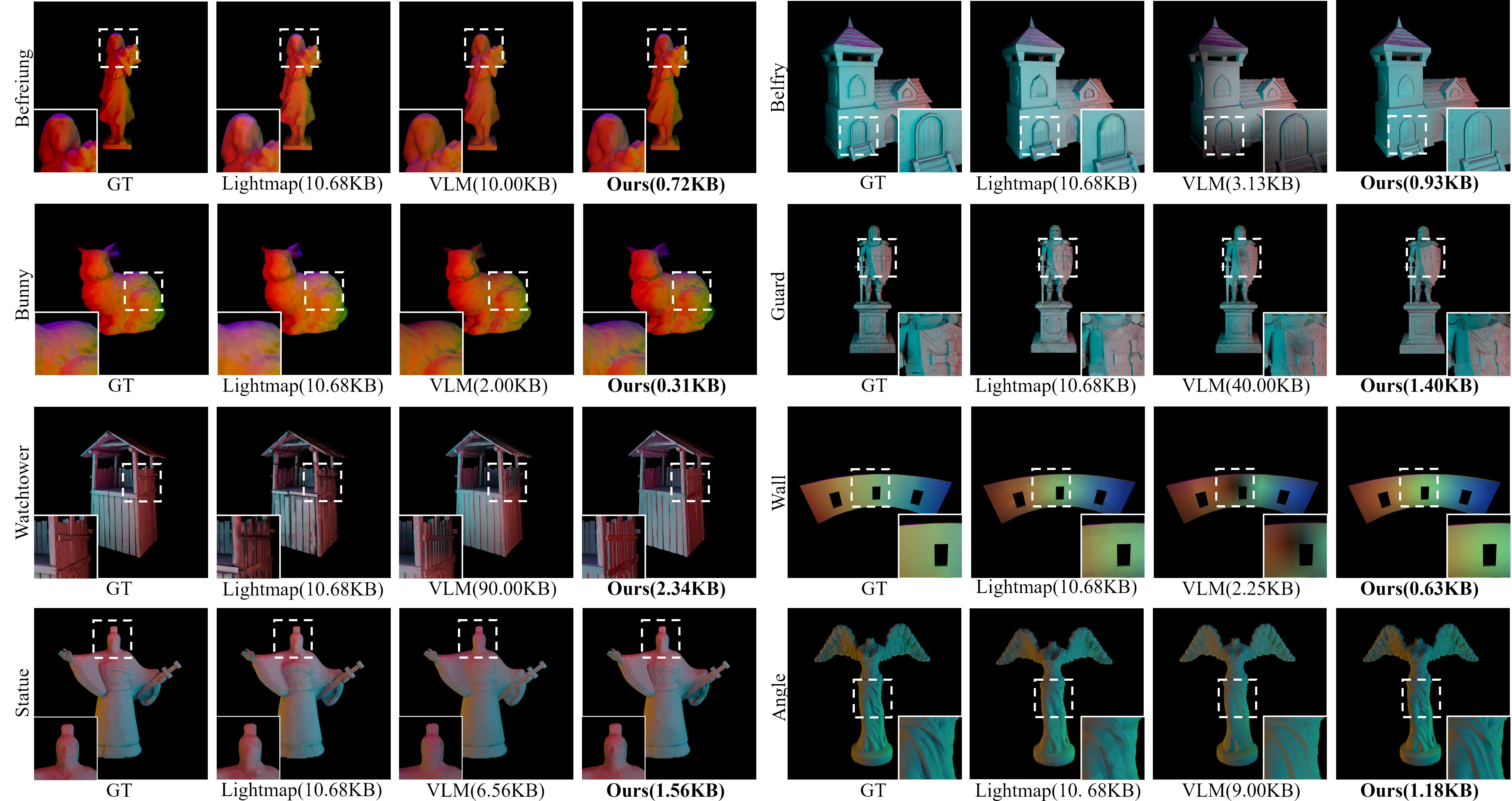}
  \caption{The results of single mesh. Each mesh was exposed to over three light sources. Our work was tested on various types of models with different characteristics, such as polygon count and the presence of occlusions, consistently achieving satisfactory global illumination with minimal memory usage.}
  \label{fig:singlemeshvisual}
\end{figure*}

\textit{Results of single mesh.} 
We tested various individual models with different complexities and types under complex lighting conditions. The quantitative results are presented in Table \ref{tab:singleresult}, while the corresponding visual effects are shown in Figure \ref{fig:singlemeshvisual}.

\begin{table}[h]
\caption{Quantitative Results of mRMSE. Tests were conducted on models of varying complexity under complex lighting, comparing our work (Probes: 20, 30) with 6x6 ASTC-compressed Directional Lightmap (Resolutions: 64, 128, 256) and two different VLM densities (Mesh-surrounding probe counts only, with memory usage shown in parentheses in the table). All memory usage in the table is measured in kilobytes (KB).}
\label{tab:singleresult}
\resizebox{0.48\textwidth}{!}{
    \begin{tabular}{l|ccc|cc|cc}
    \toprule
    $mRMSE$$\downarrow$ & \multicolumn{3}{c|}{Lightmap } & \multicolumn{2}{c|}{VLM } & \multicolumn{2}{c}{Ours} \\ 
    Mesh | Mem. & 2.67KB & 10.68KB & 42.72KB & - & - & 0.63KB & 0.94KB\\
    \midrule
    Befreiung & 0.045 & 0.041 & 0.039 & 0.041(2.0KB) & 0.032(10KB) & 0.027 & 0.025\\
    Bunny    & 0.047   & 0.046   & 0.045   & 0.043(2.0KB)   & 0.034(16KB)   & 0.031  & 0.029\\
    Watchtower    & 0.055   & 0.038   & 0.030   & 0.044(4.0KB)   & 0.041(9KB)   & 0.041   & 0.039\\
    Statue    & 0.033   & 0.031   & 0.029   & 0.027(6.6KB)   & 0.021(16KB)   & 0.021   & 0.021\\
    Townhall  & 0.052   & 0.040   & 0.034   & 0.072(3.1KB)   & 0.047(9KB)   & 0.038   & 0.036\\
    Guard    & 0.072   & 0.058   & 0.047   & 0.060(4.0KB)   & 0.047(9KB)   & 0.053   & 0.050\\    
    Wall    & 0.035   & 0.035   & 0.034   & 0.049(2.3KB)   & 0.053(15KB)   & 0.035   & 0.031\\
    Angle    & 0.041   & 0.035   & 0.034   & 0.032(3.8KB)   & 0.027(9KB)   & 0.021   & 0.020\\
    \midrule
    AVG. $\downarrow$ & 0.048 & 0.041 & 0.037 & 0.046(3.5KB) & 0.038(12KB) &0.033 & 0.031\\
   
    \bottomrule
    \end{tabular}
}
\end{table}

\begin{figure*}[h]
  \includegraphics[width=\textwidth]{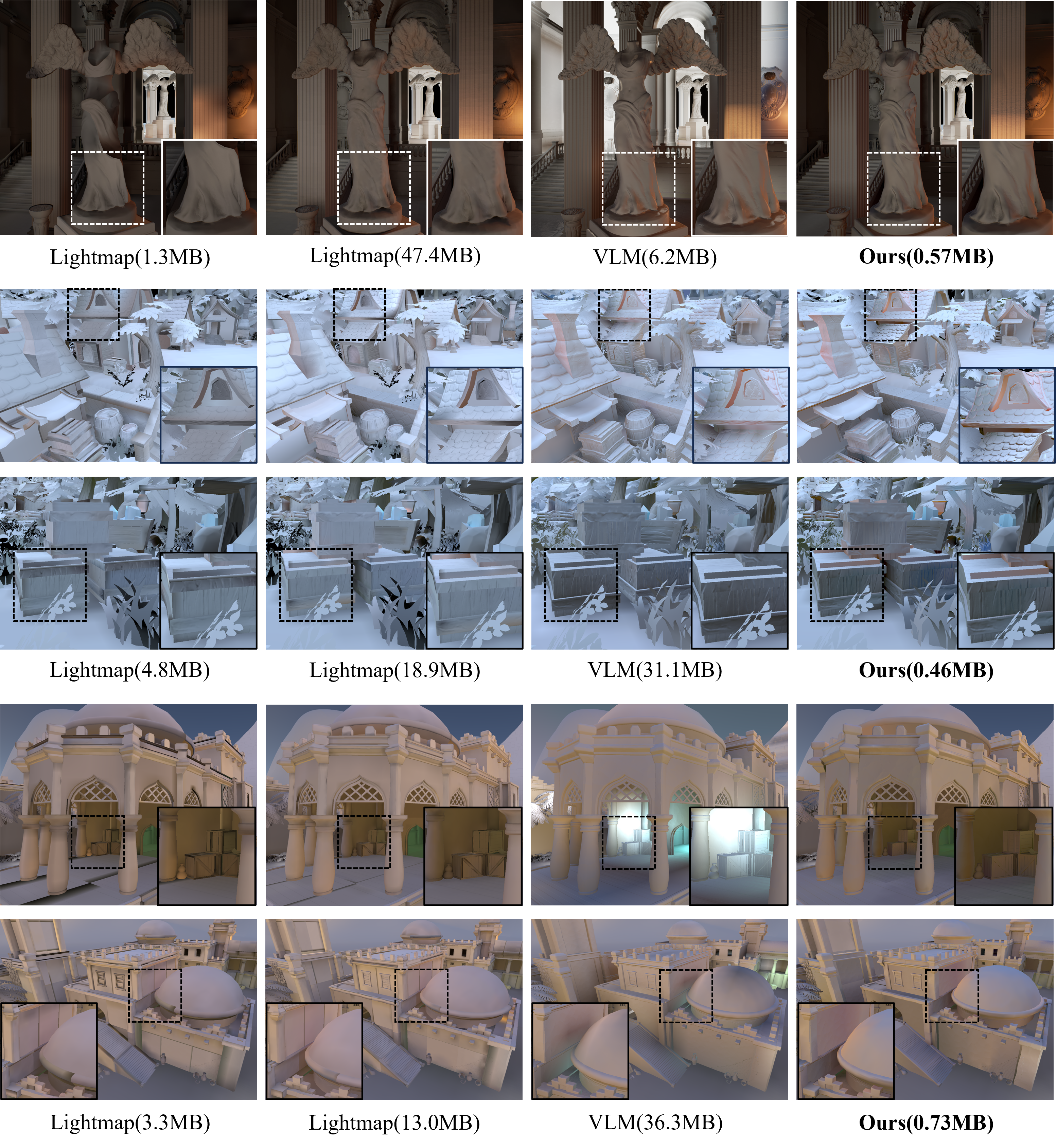}
  \caption{The results of entire scenes. We conducted tests in Unreal Engine's Sun Temple, Summer HandPainted Environment, and Stylized Egypt. Compared to other mainstream methods, our work achieves fully usable visual effects with minimal memory usage.}
  \label{fig:scenesvisual}
\end{figure*}

\textit{Results of entire scene.}
We applied our work to entire scenes (all static objects). We present a visual comparison with mainstream industry methods in Figure \ref{fig:scenesvisual}.

\subsection{Performance evaluation}

Directional Lightmap (Lightmap) \cite{lightmap} achieves top-tier runtime performance among existing methods due to its reliance solely on UV coordinates for texture lookups. Using Snapdragon Profiler, we tested our approach in the Sun Temple scene (Table \ref{tab:performance1}) and measured frame rates on four mobile devices (Table \ref{tab:performanceresult}) in a large game map (2km $\times$ 2km). By comparison, our method alleviates the burden on the fragment shader, performing efficiently on low-end platforms.

\begin{table}
\caption{The performance statistics on Samsung Galaxy S21 with the Qualcomm Snapdragon 888 and Adreno 660 GPU. We analyzed the number of texture samples for each vertex (V.) and fragment (F.), as well as the shading time consumption ratio for vertices and fragments. We effectively reduced the workload on the fragment shader, achieving ideal results in memory read bandwidth (BW.) per frame (Frm.) and basepass (B.) time.}
\label{tab:performance1}
\resizebox{0.48\textwidth}{!}{
    \begin{tabular}{l|cc|cc|c|c}
    \toprule
        & \multicolumn{2}{c|}{Tex. Samps. $\downarrow$} & \multicolumn{2}{c|}{Shading \%} & \multicolumn{1}{c|}{Mem. Read BW. $\downarrow$} & \multicolumn{1}{c}{B. Time $\downarrow$}\\
        & Tex./V. & Tex./F.  & V.(\%) & F.(\%) & Mem./Frm. (MB) & Time (ms)  \\
    \midrule
        VLM  & 0.613 & 12.20 & 1.91 & 98.09 & 43.2 & 6.83\\
        Lightmap  & 0.657 & 5.14 & 3.31 & 96.69 & 43.1 & 5.12\\
        Ours (LOD 1) & 1.239 & \textbf{3.08} & 5.22 & 94.78 &  \textbf{42.7} &  \textbf{4.95}\\
        Ours (LOD 0) & 1.315 & \textbf{3.08} & 6.34 & 93.66 &  \textbf{42.7} & 5.66\\ 
    \bottomrule
    \end{tabular}
}
\end{table}

\begin{table}
\caption{The performance test on mobile platforms. For each device, the frame rate (FPS) represents the average value across multiple identical units. Our work achieves performance comparable to that of the Lightmap, enabling good support for low-end devices.}
\label{tab:performanceresult}
\resizebox{0.45\textwidth}{!}{
    \begin{tabular}{l|c|c|c|c|c}
    \toprule
        & \multicolumn{1}{c|}{FindX3Pro} & \multicolumn{1}{c|}{Mi 6} & \multicolumn{1}{c|}{Pixel6Pro} & \multicolumn{1}{c|}{GalaxyS9+} & \multicolumn{1}{c}{Avg. $\uparrow$}\\
    \midrule
        Lightmap  & 55.96 & 18.70 & 51.36 & 15.03 & 35.26\\
        Ours & 55.67 & 16.46 & 52.36 & 22.13 &  \textbf{36.65}\\
    \bottomrule
    \end{tabular}
}
\end{table}

\subsection{Ablation study}

\begin{figure}[h]
  \centering
  \includegraphics[width=\linewidth]{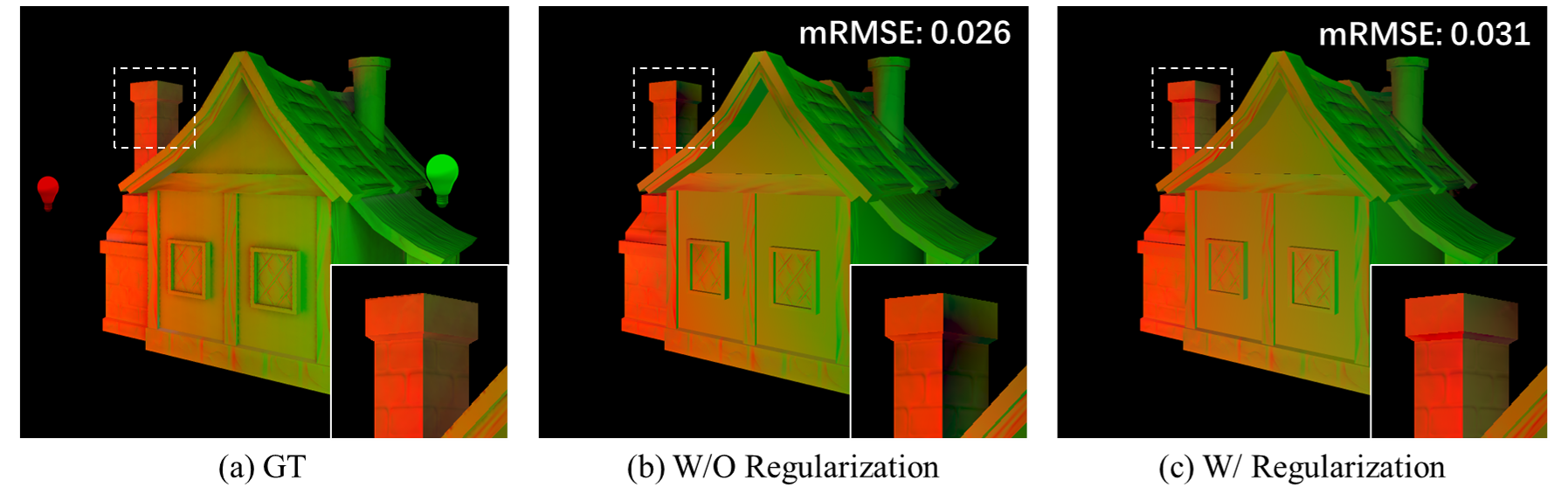}
  \caption{The Impact of Gradient Regularization (Forge with 40 probes, 1.25 KB). Without regularization (b), lower loss results in noticeable unsmooth visual artifacts.}
  \label{fig:reg}
\end{figure}

\textit{Gradient regularization.}
We validated the impact of the regularization term in Eqn. (7). As shown in Figure \ref{fig:reg}, the baking method without the regularization term focuses solely on the numerical accuracy of the illumination, disregarding visual smoothness. While the error is numerically lower without regularization, it introduces noticeable visual artifacts.

\begin{figure}[h]
  \centering
  \includegraphics[width=\linewidth]{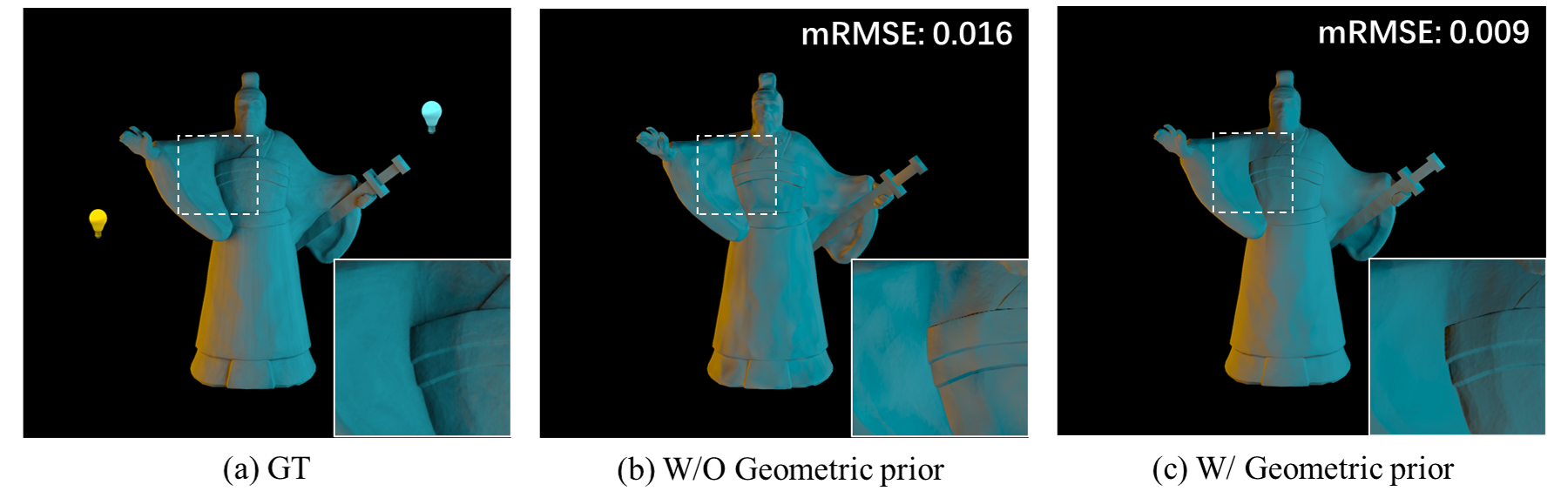}
  \caption{The Impact of Geometric Prior (Statue with 50 probes, 1.56KB). Without geometric prior (b), light leakage remains an issue after 5000 epochs. By initializing probes with the geometric prior and then optimizing (c), satisfactory associations are achieved in just 400 epochs.}
  \label{fig:geo}
\end{figure}

\textit{Geometric prior.} We show the impact of the geometric prior mentioned in Section 3.4 in Figure \ref{fig:geo}. By initializing probes with geometric priors and further optimizing through gradient descent, light leakage is effectively reduced, and the optimization time is significantly shortened. This result also demonstrates the effectiveness of our probe distribution strategy.


\subsection{Discussion and limitations}
\hspace{0.6\parindent}
\textit{No UV mapping requirement.} 
Our work relies on geometric vertices, removing the need for UV mapping. Traditional workflows require precise, non-overlapping UV maps, which consume storage, waste memory due to gaps, and are time-intensive to create. Our approach saves both storage space and production time related to UV mapping.

\begin{figure}[h]
  \centering
  \includegraphics[width=0.8\linewidth]{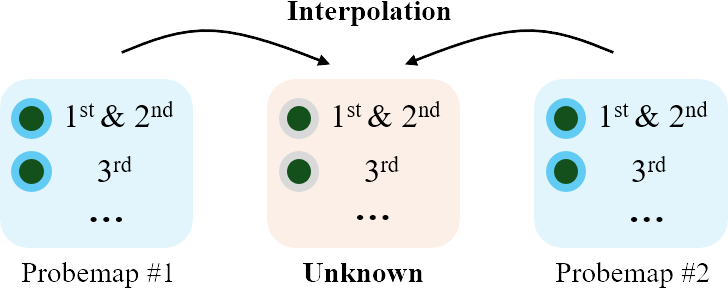}
  \caption{The probemap at a given moment is derived through interpolation between the known probemaps of adjacent time points.}
  \label{fig:probemapinter}
\end{figure}

\begin{figure}[h]
  \centering
  \includegraphics[width=\linewidth]{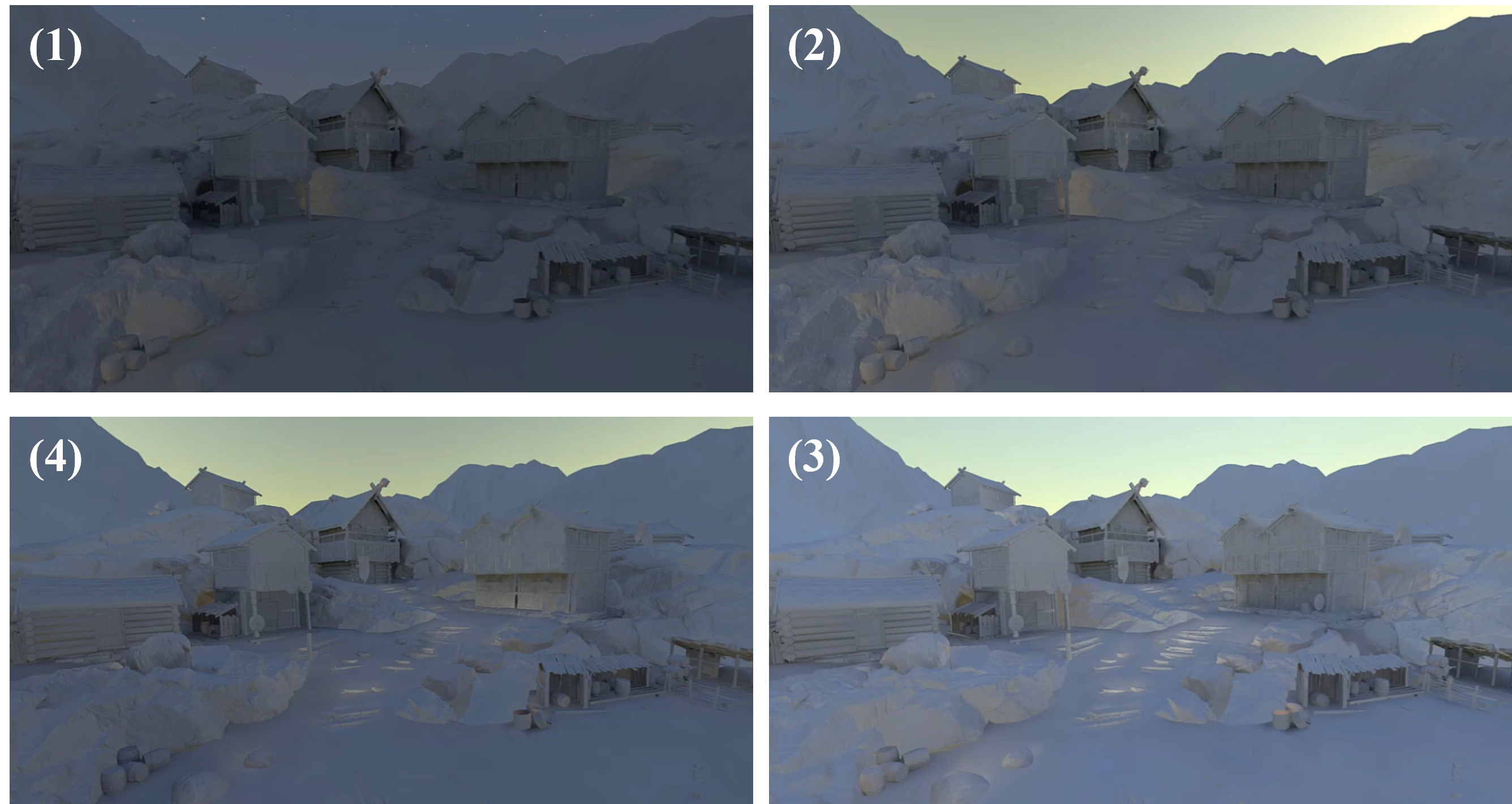}
  \caption{Time of the Day in LightingTutorial Scene. Our work requires only 4MB of memory (0.5MB $\times$ 8) to realize the day-night global illumination effect. In contrast, the traditional lightmap method requires 18MB of memory usage for a single time point at comparable quality.}
  \label{fig:day2night}
\end{figure}

\textit{Time of Day support.} 
Our work enables straightforward implementation of Time of Day (TOD) effects by interpolating between probemaps, as shown in Figure \ref{fig:probemapinter}. This process can be efficiently conducted in the compute shader for the entire scene (or a large region in current use) due to its small texture size, ensuring that it does not impose any burden or shader complexity on per object drawing. For example, we achieved global illumination for the entire scene's day-night cycle by interpolating among 8 probemaps (corresponding to 8 different time points), as shown in Figure \ref{fig:day2night}.

\textit{Visual quality across different meshes.}
The visual quality of our method is determined by the number of probes and the vertex density of the mesh. Unlike existing approaches, our method performs better on complex models with denser meshes. For extremely simple meshes, higher accuracy can be achieved by applying simple subdivision operations to increase mesh density, but this will also increase the geometric memory usage. In our engineering practice, models satisfying the vertex density criteria constitute more than 95\% of the scene and using traditional methods for extremely simple meshes is a better choice. For meshes that do not require reuse or are highly complex, we can opt for targeted optimization, such as distributing probes based on the actual lighting environment. This will effectively reduce the number of light probes while improving lighting quality.

\textit{Dynamic objects.}
Our work mainly focuses on static objects. For dynamic objects in mobile games, which are often augmented with supplementary lighting, the reliance on global illumination (GI) is less critical compared to static objects. Simpler GI solutions are typically more suitable for low-end platforms, such as Indirect Lighting Cache (ILC), which conduct an interpolation for the whole component. As for extending our work to dynamic objects, we consider interpolating SH values from the surrounding environment to be a viable approach.

\textit{Potential LOD popping.}
Given that visual requirements on low-end platforms are generally less stringent, and the impact of global illumination (GI) transitions is less pronounced than that of mesh level of detail (LOD) changes, our work simply applies probe distribution to each level of the mesh, which may lead to potential popping phenomena. We consider simplifying the probe distribution from LOD0 to be a valuable task, as it can save optimization time and ensure smoother transitions. We will investigate this further in the future.

\textit{High-frequency detail limitation.} 
Due to the constraints of spherical harmonics (SH) and the vertex-probe structure, it is difficult for our work to represent high-frequency details such as sharp shadows. We explored using a lightweight neural network to address this issue but found it challenging to balance performance and quality. Considering that this limitation is not critical for low-end platforms, we did not conduct further exploration and leave this problem for future work.

\begin{figure*}[p]
  \includegraphics[width=\textwidth]{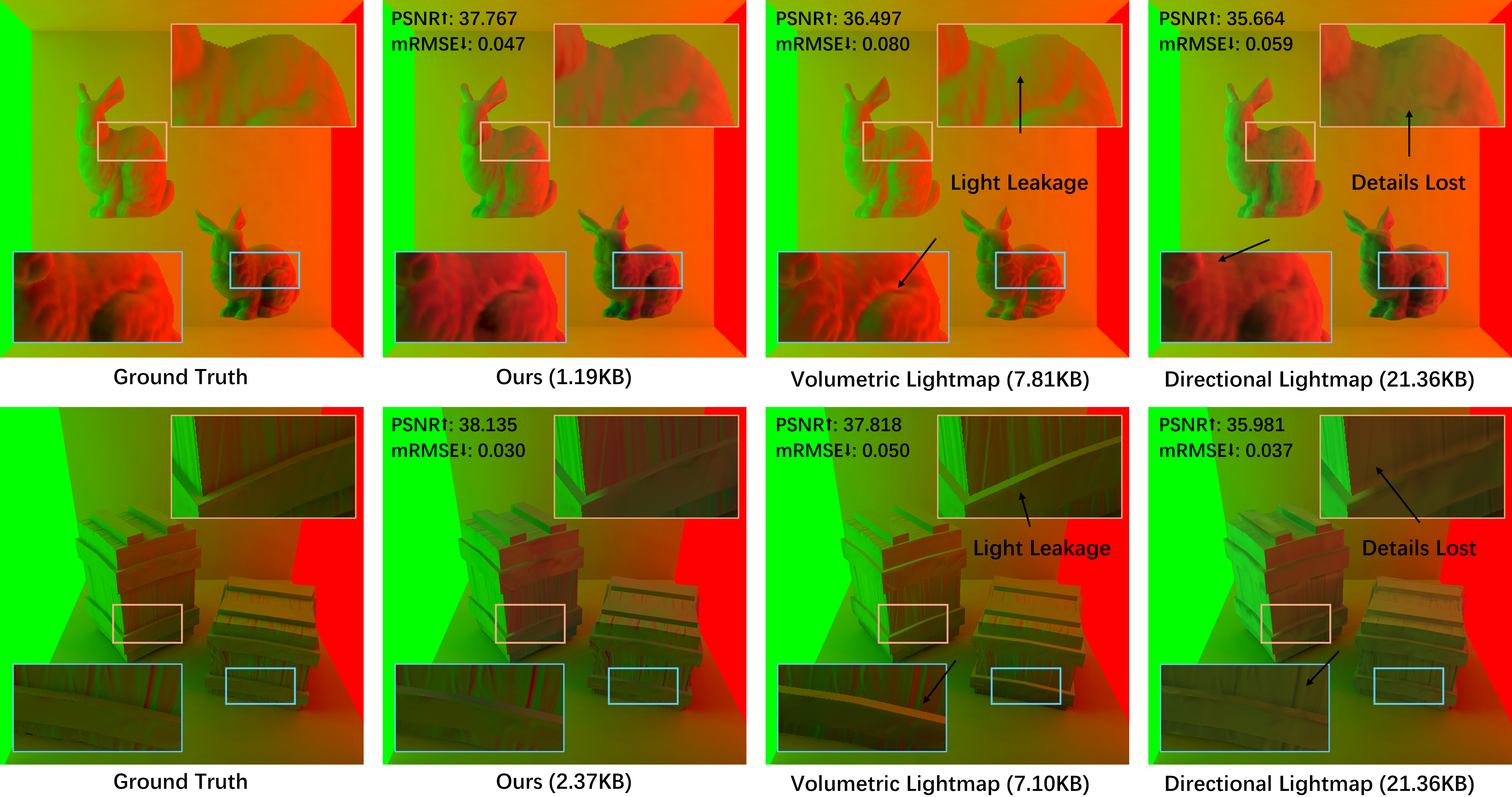}
  \caption{The results of multiple objects in a Cornell Box with the left and right walls as emissive light sources. Our work effectively reduces light leakage and preserves more details. Compared to PSNR, mRMSE is more sensitive to light leakage due to the hemispherical lighting information of the whole object.}
  \label{fig:cornellbox}
\end{figure*}

\section{CONCLUSION}

In this paper, we present a new static global illumination solution tailored for low-end platforms. We introduce a novel illumination baking model that achieves high-precision fitting for effective illumination and formulate probe distribution as an optimization problem, enabling inverse probe placement in local space. Thanks to our efficient illumination reconstruction method and fitting strategy, our method achieves high performance while maintaining precision and supporting Level of Detail (LOD). Moreover, it delivers highly competitive lighting effects with an extremely low memory usage and effectively reduces sampling in fragment shader. This makes our work highly suitable for addressing the static global illumination requirements of low-end platforms.

The demand for high-quality virtual scenes on mobile devices is rapidly increasing. However, most existing works focus on PC platforms, aiming for higher-quality lighting effects, leaving very limited options for low-end platforms. Therefore, designing pipelines around high performance and low memory usage is both valuable and challenging. We hope our work can inspire further exploration in the lightweight optimization.

\begin{acks}
We thank the reviewers for the valuable comments. We thank the Activision Research, whose previous work greatly inspired us. This work was supported in part by the National Key R\&D Program of China (No. 2022YFB3303400).

\end{acks}

\bibliographystyle{ACM-Reference-Format}
\bibliography{reference.bib}

\end{document}